# A Study of Computer-Based Simulations for Nano-Systems and their types

*Tamal Sarkar[1], Samir Chandra Das[2], Ardhendu Mandal[3]*
**North Bengal University**
**P. O., N.B.U, Siliguri-734013, India**
[1][University Science Instrumentation Centre (USIC) , usic@nbu.ac.in]
[2][USIC, samirdasju@yahoo.co.in]
[3][Computer Science and Applications (CSA), am.csa.nbu@gmail.com]

**ABSTRACT**
In most of the cases, the experimental study of Nanotechnology involves high cost for Laboratory set-up and the experimentation processes were also slow. So, one cannot rely on experimental nanotechnology alone. As such, the Computer-Based molecular simulations and modeling are one of the foundations of computational nanotechnology. The computer based modeling and simulations were also referred as computational experimentations. In real experiments, the investigator doesn't have full control over the experiment. But, in Computational experimentation the investigator have full control over the experiment. The accuracy of such Computational nano-technology based experiment generally depends on the accuracy of the following things: Intermolecular interaction, Numerical models and Simulation schemes used. Once the accuracy of the Computational Scheme is guaranteed one can use that to investigate various nonlinear interactions whose results are completely unexpected and unforeseen. Apart from it, numerical modeling and computer based simulations also help to understand the theoretical part of the nano-science involved in the nano-system. They allow us to develop useful analytic and predictive models. In this paper, a brief study of Computer-Based-Simulation techniques as well as some Experimental result obtained using it were given.

## INTRODUCTION

The computer based simulation techniques [1] are widely used for computational nano-technology. The frequently used simulation approaches are Monte Carlo (MC) and Molecular Dynamics (MD) methods. The other various simulation methods come from these two basic methods. Apart from it, the optimization techniques were also needed in MC and MD. The computer- based simulation methods, developed for nano-systems, generally consist of a computational procedure performed of few atoms or molecules confined in a small geometrical space. This geometrical space in which the simulation is performed is termed as *cell*. In the subsequent section, a brief classification of simulation methods based on Accuracy, Computational Time etc were given. After it, classifications of optimizations in molecular simulations were discussed. In next section after it, we have discussed the result of computer -based simulation.

## MONTE CARLO AND MOLECULAR DYNAMICS SIMULATION METHODS [2, 3, 6]

The two basic used simulation approaches are Monte Carlo (MC) and Molecular Dynamics (MD) methods. All the other various simulation methods come from these two basic methods. A brief over-view with areas of application of the both are discussed below. These concepts are essentially required to understand the methodology of classification of Computer-Based Simulation methods based on accuracy and time-complexity.

MC method uses random numbers to perform calculations. There are many areas of application of MC Methods including nano-material. Some important areas where we apply MC method are: - (i) Estimation of large-dimensional integrals (ii) Generating thermodynamic ensembles in order to compute thermal averages of physical equilibrium quantities of interest and simulation of non-equilibrium phenomena such as growth and (ii) Computation of distribution functions out of equilibrium known as *Kinetic Monte Carlo [5]*.

MD deals with predicting the trajectories of atoms subject to their mutual interactions and eventually an external potential. Some important areas of application of MD are: - (i) Computation of transport properties such as response functions, viscosity, elastic moduli and thermal conductivity (ii) Thermodynamic properties such as total energy and heat capacity and (iii) Dynamical properties such as phonon spectra.

## CLASSIFICATION OF SIMULATION METHODS BASED ON ACCURACY AND COMPUTATIONAL TIME

Computer based methods used for simulation of various properties of nano scale systems differ in their level of accuracy and time-complexity to perform such calculations. Based on it, the required time scale for these methods can be from tens of picoseconds (e.g. ab initio MD calculations) to few microseconds or more (classical molecular dynamics simulation). There are also methods which require very long computational time such as cluster growth and may require super computers to achieve fast results.

Based on these facts we may classify the methods into following groups (i) Methods with highest degree of accuracy (ii) Methods with second highest degree of accuracy (iii) Semi-empirical method and (iv) Stochastic method (shown in Table 1) **[4, 5]**.

| TABLE-1 | | | | |
|---|---|---|---|---|
| **Methods** | **Purpose** | **Input** | **Output** | **Examples** |
| Methods with highest degree of accuracy | Investigation of both electronic and atomic ground state, optical and magnetic properties of weakly interacting and also strongly interacting correlated systems | Atomic species, coordinate, system's symmetry, interaction parameter | Total energy, excitation energy and spin densities, force on atoms | Ab initio methods for electronic structure calculations of correlated systems, Quantum MC, Quantum Chemistry and Many body |
| Methods with second highest degree of accuracy | Accurate calculations of ground state structure by local optimization; Calculation of mechanical, magnetic and optical properties of small clusters and perfect crystals of weakly interacting electron systems, estimation of reaction barrier and paths | Atomic species and their coordinates, symmetry of the structure, pseudo-potential or Hamiltonian parameters for the species considered | Total energy, charge and spin densities, forces on atoms, electron energy eigen values, capability of doing Molecular Dynamics, vibrational-modes and phonon spectrum | Ab initio methods for normal Fermi liquid systems based on either Hartree Fock or Density Functional Theories |
| Semi-empirical method | Search for ground state structure by GA, Simulated Annealing (SA) or local optimization if a good guess for the structure is known; simulation of growth or some reaction mechanisms; calculation of response functions | Atomic species, their coordinates; parameters of the inter-particle potential, temperature and parameters of the thermostat or other thermo-dynamic variables. | Output of Tight-Binding(TB): Total energy, charge and spin densities, force on atoms, particle trajectories, phonon calculation; mechanical magnetic and optical properties of clusters and crystals | Semi-empirical methods of large systems or long time scale. TB or LCAO molecular dynamics based on classical potential or force-field |
| Stochastic method | Investigation of long timescale non-equilibrium phenomena such as transport, growth, diffusion, annealing, reaction mechanisms and also calculation of equilibrium quantities and thermodynamic properties | Parameters of the inter-particle potential, temperature and parameters of the thermostat or other thermodynamic variables | Statistics of several quantities such as energy, magnetization, atomic displacements | MC walk towards equilibrium, kinetic or dynamical MC (growth and other non-equilibrium phenomena) |

## CLASSIFICATION OF OPTIMIZATIONS IN MOLECULAR SIMULATIONS

*Optimization* [4, 7] in any numerical computation is to look for the best value of a multivariate function among the many numbers generated during the computation. These are common in molecular simulations where a system consisting of assembly of N particles is described in a 6N-dimensional phase space. In such cases, a cost (here energy) is associated with every configuration and objective is to seek sets of data points that maximize (or minimize) the objective function. There are two kind of optimizations required in every molecular simulation (i) Local optimization (ii) Global optimization (Fig. 1)

Some examples in each type of optimizations are discussed below:-
Local Optimization Methods: (i) Steepest Descent Methods (ii) Damped Newtonian Dynamics (iii) Conjugate Gradient Method (iv) Quasi- Newton Method
Global Optimization Methods: (i) Simulated Annealing Method (ii) Genetic Algorithms

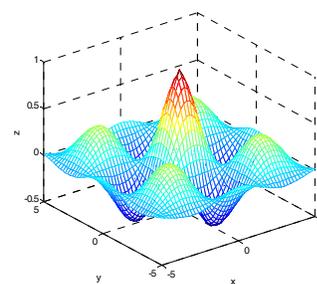

*Fig.1. Local and Global Maxima*

## Local Optimization Methods

### Steepest Descent Method (SDM)

This is the simplest optimization method. In this, one has to move along force applied on the particle. This method certainly works because at each step the potential energy is lowered, but its convergence is often slow. Even a situation may arrive where a particle will keep oscillating around the anisotropic valley and may take may step for reaching the minimum.

The simple iterative formula to implement SDM is $x(t + \Delta t) = x(t) + \lambda f[x(t)]$ where $f(x) = -\nabla E(x)$ the force is. In its discretized form, it can be written as $x(i+1) = x(i) + \lambda f[x(i)]$.

### Damped Newtonian Dynamics Method

This is faster than SDM. Here we consider, the algorithms
$x(i+1) = [2x(i) + \lambda f(x(i)) - x(i-1)(1-\mu)]/(1+\mu)$

### Conjugate Gradients Methods (CGM)

It is more intelligent method than SDM and Damped Newtonian Dynamics Method. CGM will find the minimum of the anisotropic valley faster than the SDM. In principle, it finds the minimum of a quadratic function in N dimensions in exactly N line minimization steps. If one is near t he minimum, the convergence to the minimum is almost as fast as the SDM.

### Quasi- Newton Methods

Newton's method: It is a method to find the zero of a function f in one dimension. The expression for the iterative scheme is as follows:

$x(i+1) = x(i) - f(x((i)))/f'(x(i))$

## GLOBAL OPTIMIZATION METHODS

### Genetic Algorithm (GA)

It is a biological evolutionary process in intelligent search, machine learning and optimization problems. This algorithm is basically search algorithm based on the machines of natural selection and nature genetics shown in *Fig 2. It is the balance between efficiency and efficacy necessary for survival in many different environments.* In order for *GA* to surpass their more traditional in the quest for robustness, GA must differ in some fundamental ways. GA are different from more normal optimization and search procedures in four ways:

i. GA work with a coding of the parameter set, not the parameters themselves.

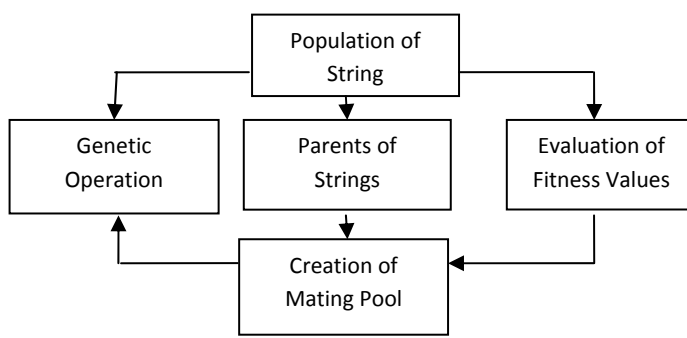

*Fig.2. Flow chart for process of a GA global optimization of a molecular simulation*

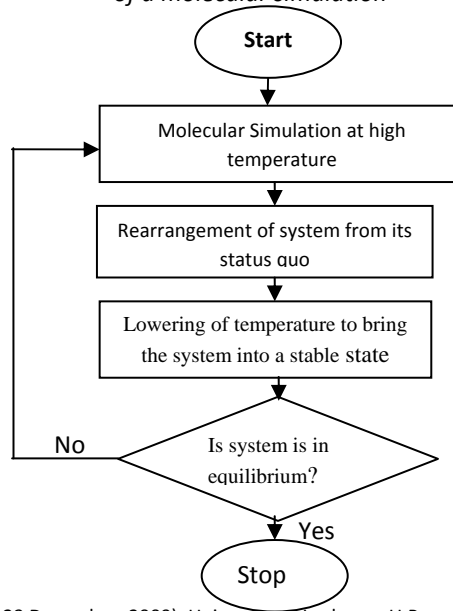

*Fig.3. Flow chart for process of a simulated annealing global optimization of a molecular simulation*

ii. GA search from a population of point not a single point.
iii. GA use pay off (objective function) not derivatives or other auxiliary knowledge.
iv. GA use probabilistic transition rules, not deterministic rules.

*Simulated Annealing Method*

This method was first proposed by Kirkpatrick et al. this method was chosen for large scale optimization problem in which required global minima is hidden between many local minima. Simulated annealing distinguishes between the different local optima in a same way analogous to the physical process of annealing. The flowchart for simulated annealing to go through for global optimization is shown in Fig. 3.

## COMPUTER BASED EXPERIMENT AND RESULTS OBTAINED USING GA

Present work declares that a soft computing tool, GA is used to get the optimized system parameters of GaAs QW for a desired high frequency response characterized by a cutoff frequency ($f_{3dB}$). By the present work, a model for optimized system parameters of GaAs, QW is obtained for a high frequency under hot electron condition. In GA, a fitness function is the main criteria for reproduction. The fitness values are used to favor high fitness individuals over low fitness individuals to take part in the process of reproduction. In the present application of GA, we find the $f_{3dB}$ for a semiconductor quantum structure for its different system parameters. So we find the variation of $f_{3dB}$ (cut off frequency at which the mobility falls to **0.707** of its low frequency value) and one particular parameter of the system where the other parameters are optimized by the GA. By taking the other parameters in one form, we can be able to find the fitness values. These fitness values are converted to binary form and then proceed for further GA operation. After reproduction, simple crossover may proceed in the following steps. First, members of newly reproduced strings in the mating pool are mated at random. Second, each pair of strings undergoes crossing over as follows: - An integer position t along the strings is selected uniformly at random between 1 and the string length less one [1, P - 1]. Two new strings are created by swapping all characters between positions (t + 1) and P inclusively. For example consider two strings X and Y.

|

X = 10011|1011

Y = 00110|0010

|

Suppose in choosing a random number between 1 and 8 (9 - 1), as P = 9 as indicated in the above by dotted line. Then the result of cross over which produces two new strings indicated by $X_1$ and $Y_1$ (say).

$X_1$ = 100110010

$Y_1$ = 001101011

**TABLE 2:** Comparison of optimized system parameters for frequency = 300GHz and $F_0$= 0.75×10$^5$ V/m of GaAs.

| ac mobility $\mu_{ac}$ (m$^2$/V.s) | $T_L$ (K) | $T_e$ (K) | $n_{2D}$ (10$^5$ m$^{-2}$) | $L_z$ (nm) |
|---|---|---|---|---|
| 1.61 | 77 | 335 | 9.0 | 95 |
| 1.83 | 100 | 170 | 5.9 | 90 |
| 1.91 | 125 | 215 | 7.0 | 95 |
| 1.60 | 150 | 190 | 5.0 | 110 |
| 1.98 | 175 | 300 | 10.0 | 120 |
| 1.78 | 200 | 275 | 8.0 | 120 |
| 1.65 | 225 | 285 | 8.0 | 95 |
| 1.56 | 250 | 305 | 8.0 | 90 |
| 1.66 | 275 | 350 | 9.0 | 115 |
| 1.37 | 300 | 345 | 7.0 | 110 |

[$T_L$= Lattice temperature, $T_e$= Electron temperature, $n_{2D}$= Carrier concentration, $L_z$= Channel length]

The mutation operator plays a secondary role in the simple GA. We simply note that the frequency of mutation to obtain good results in empirical GA studies is on the order of one mutation per thousand bit transfers.

## NUMERICAL RESULTS

In this optimization technique, we consider a square quantum well of GaAs and electron temperature model is employed for numerical calculations with the following system parameters : *Electron effective mass m\* = 0.3735 x $10^{-31}$kg, Static dielectric constant $K_s$ = 13.88, optic dielectric constant $K_α$ = 11.34, LO phonon angular frequency $ω_0$ = 5.37x $10^{13}$ rad $s^{-1}$, longitudinal elastic constant $C_L$ = 14.03 x $10^{10}Nm^{-2}$, acoustic deformation potential constant.$E_1$= 17.6x $10^{-11}$ J, background ionized impurity concentration $n_{bi}$ = 6.0 x $10^{21}m^{-3}$, longitudinal acoustic velocity, $u_1$ = 5.22 x $10ms^{-1}$*.

The frequency of the applied is varied and at each frequency the optimum parameters are determined by employing the Genetic algorithm for a particular dc biasing field. For a particular dc biasing field, it is possible to predict the optimum values of the system parameters like electron temperature, channel width, carrier concentration, for the realization of desired high frequency response and ac mobility[8-11]. The optimized parameters will obviously save the search time for the technologists involved in the fabrication of the devices.

We have computed the optimized parameters using GA to get desired millimeter and sub-millimeter wave response characteristics. The application of GA in the device parameter optimization for getting the best performance is a new area of research. The optimized system parameters presented here predict the better performance of GaAs QWs in the microwave and millimeter wave regime and can be used to analyse the experimental data when they appear in the literature. Artificial neural network and genetic algorithm based computations are performed with the parameter values of $In_{0.53}Ga_{0.47}As$ for the dc biasing field $F_0$ of 0.75×$10^5$ V/m and the frequency of the applied ac field of 300GHz.

TABLE 3: Optimized parameters for dc bias field $F_0$ = 1.0 x $10^5$ v/m

| Frequency (GHz) | $n_{2D}$ ($10^{15}$ $m^{-2}$) | $L_z$ (nm) | $T_L$ (k) |
|---|---|---|---|
| 110 | 10.0 | 108 | 85 |
| 135 | 8.0 | 118 | 115 |
| 160 | 9.0 | 115 | 182 |
| 180 | 10.0 | 122 | 280 |
| 210 | 9.0 | 103 | 285 |
| 230 | 6.0 | 88 | 237 |
| 250 | 6.0 | 85 | 290 |

[$n_{2D}$= Carrier Concentration, $L_z$=Channel Width, $T_L$( K) = Lattice Temperature]

Optimized system parameters for the frequency 300GHz of the applied small-signal electric field and the dc biasing field $F_0$ of 0.75×$10^5$ V/m are given in Table 2, which reveal that for the desired cutoff frequency at a particular dc biasing field it is possible to predict the optimum values of the system parameters like carrier concentration, channel length and electron temperature for realizing a particular high frequency response characterized by a cutoff frequency. It is found that at the lattice temperature $T_L$ of 200K, the optimized system parameters are: electron temperature $T_e$=275K, 2D carrier concentration $n_{2D}$= 8×$10^5m^{-2}$ and the channel length $L_z$= 120 nm. The corresponding parameters obtained by GA are obtained placed in the Table 3. Comparison recalls that the ANN results are also equally suitable device GA based results.

## Conclusion

This paper demonstrates the various processes through which computer based simulations and optimizations are used in various field starting from classical to nano-levels. The same will also help the reader to choose the best method suitable for a particular application of their interest. To give a better view we have given one example of computational technique that we have used for nano device simulation from one of our experimental work.